\documentstyle[epsfig]{aipproc2}

\begin{document}
\title{Relic gravitons from non-singular \\
string cosmologies}

\author{Cyril Cartier}
\address{Centre for Theoretical Physics, University of Sussex,
\\ Falmer, Brighton BN1~9QH, U.K.}

\maketitle

%------------------------------------------------------------------

\begin{abstract}
In the context of the pre-Big Bang scenario of string cosmology,
we propose a modified equation for the evolution of the tensor perturbations,
which includes the full contribution of possible higher-order curvature and 
coupling corrections required to regularise the background evolution. 
We then discuss the high-frequency branch of the spectrum of primordial 
gravitons. Preliminary results suggest that the slope of the string 
branch of the spectrum is too steep to allow for the cosmological background 
of relic gravitons to match the frequency range probed by 
planned experiments such as interferometric detectors.
\end{abstract}

\section*{Introduction}
Recent studies, including strong coupling corrections to the string effective
action, have provided the pre-Big Bang scenario 
\cite{Gasperini:1991rv,Gasperini:1993em}
with a number of promising models for the transition from the growing
to the decreasing curvature regime
\cite{Brustein:1998cv,Foffa:1999dv,Brustein:1999yq,Cartier:1999vk}.
However, a crucial question remains: is it possible to test this scenario
and, in particular, to distinguish it from other candidate models which 
attempt to describe the very early universe?

The answer to this essential question is in principle in the affirmative, 
since the pre-Big Bang scenario incorporates a dilaton field coupled to the 
metric, which modifies the evolution of tensor perturbations. 
It is well known, indeed, that the transition from an inflationary period to
the FRW radiation-dominated era is associated with
the production of a background of relic gravitational waves. Such
primordial signals decoupled from matter soon after the Planck
era and, unlike electromagnetic radiation which underwent a
complicated history until recombination, gravitons have been transmitted
scarcely without interaction down to our present epoch. As a
consequence, their present spectrum should be a faithful portrait of the
very early universe, thus opening a window for the observation of
processes occurring near the Planck scale, and for
discriminating among various theories of high-energy models of
cosmology and of unified interactions.
With respect to the standard inflationary scenario, the
amplification of tensor perturbations in the pre-Big Bang scenario
is strongly enhanced in the limit of large frequencies
\cite{Gasperini:1992pa,Gasperini:1993dp},  and the amplitude of the spectrum
is normalised so as to match the string scale at the high-frequency end point
of the spectrum \cite{Brustein:1995ah,Brustein:1997ut}.
As a result, it has been claimed \cite{Gasperini:1993em,Brustein:1995ah}
that the produced background of 
gravitational waves could be detected in the near future by various 
experiments such as the second 
(planned) generation of interferometric detectors 
\cite{Thorne:1997ut,Allen:1997ad}.

Up to now, the above extended studies of tensor perturbations have been 
mostly performed in the context of the lowest-order effective action of 
string theory. However, the singular behaviour of the tree-level background 
cosmological solutions implies that such studies are a priori inadequate to 
describe the high-frequency branch of the spectrum. Indeed short-scale modes 
are expected to leave the Hubble radius during the high-curvature regime which
has to remain bounded from above. This is a major problem for the prediction
of the amplification of these high-frequency modes,
the ones to be mostly detected (if at all).

The aim of this contribution\footnote{Contribution to CAPP~2000, 
Verbier (Switzerland), July 2000.
To appear in the Proceedings (American Institute of Physics publication).} 
is to discuss the evolution and
the spectral distribution of primordial gravitons in the context of 
non-singular cosmological backgrounds, by taking into account (in the
perturbation equation) the full contribution of those corrections that
are responsible for the regularisation of the background solution.

\section*{Non-singular backgrounds}
In the pre-Big bang framework, our FRW-universe is presently understood
to result from the collision of plane waves \cite{Feinstein:2000ja} emerging
from the trivial string vacuum in the asymptotic past \cite{Buonanno:1998bi}.
From this epoch onwards, the evolution of the universe, driven by the kinetic
energy of the dilaton, undergoes an inflationary expansion with growing 
coupling and curvature. Such an evolution is usually derived by the 
$4$-dimensional low energy effective action of string theory:
\begin{equation}\label{Low}
S_{eff} = \frac{1}{16 \pi \alpha'} \int d^4x \sqrt{-g}\Bigl\{e^{-\phi}
\Bigl[ R + \phi_{,\mu} \phi^{,\mu} \Bigr] +{\cal L}_{c} \Bigr\}.
\end{equation}
When the curvature reaches a maximal scale $|H| \sim {\cal O}(
\lambda_{S}^{-1})$, where $\lambda_{S} \sim \sqrt{\alpha'}$,
the universe is expected to be smoothly connected to the FRW regime,
with a constant dilaton field. The most promising approach to the graceful 
exit problem \cite{Brustein:1994kw} suggests a cure  to the curvature and 
dilaton singularities by adding higher-order corrections to the string 
effective action (see for instance
\cite{Antoniadis:1994jc,Rey:1996ad,Gasperini:1997fu,Brustein:1998cv,Foffa:1999dv,Cartier:1999vk}),
whose sources are twofold: tree-level $\alpha'$ corrections,
resulting from the string tension expansion of the effective action
and loop corrections, arising from the more conventional loop
expansion in powers of the string coupling $g_s = e^{\phi/2}$.
The tree-level $\alpha'$ correction we will be considering
(see \cite{Cartier:1999vk} for a detailed analysis) includes the
most general form for a correction to the string action up to fourth-order
in derivatives \cite{Metsaev:1987zx}:
\begin{equation}\label{Class}
   {\cal L}_{\alpha'} = \alpha'\lambda
       e^{-\phi} \Bigl\{ c_1 R_{GB}^2 + c_2 \Bigl[R^{\mu\nu}
-g^{\mu\nu}R\Bigr] \phi_{,\mu} \phi_{,\nu}
      + c_3 \Box \phi\, \phi^{,\mu} \phi_{,\mu}  
- c_4 (\phi^{,\mu} \phi_{,\mu})^2 \Big\},
\end{equation}
where $R_{GB}^2 \equiv R^{\mu\nu\rho\sigma} R_{\mu\nu\rho\sigma} 
- 4 R^{\mu\nu}
R_{\mu\nu} + R^2$ is the well-know Gauss-Bonnet combination
ensuring the field equations will remain second order in the
fields. 
In fixing the coefficients $c_i$'s, we require that the full action reproduces
the string scattering amplitude, and thus impose the constraints
$2 c_3 = -[c_2 + 2 (c_1 + c_4)]$ and $c_1 =-1$, working in units $\alpha'=1$ 
\cite{Metsaev:1987zx}. The parameter $\lambda$ allows us to move 
between different string theories and later we will set
$\lambda_{0}=-1/4$ to agree with previous studies of the Heterotic string 
\cite{Gasperini:1997fu}. Another conventional expansion of string theory 
relies on the string coupling parameter $g_s$. 
There is as yet no definitive calculation of the full loop
expansion of string theory, and we are left to speculate on plausible terms
that will eventually make up the loop corrections. Multiplying each term
of the tree-level $\alpha'$ correction by a suitable power of the string 
coupling is the approach we will be considering here and has already met 
with some success \cite{Brustein:1998cv,Brustein:1999yq,Cartier:1999vk}.
Since the quantum loop corrections are not formally derived from
a string loop expansion, we shall allow different coefficients
$d_i$ ($e_i$) at one-loop in the string coupling (two-loop respectively),
which are not necessarily subject to the previous constraints.
The effective lagrangian density ${\cal L}_{c}$ in Eq.~(\ref{Low}) is thus
 a sum of the tree-level $\alpha'$ and loop corrections,
 ${\cal L}_{c} = {\cal L}_{\alpha'} + {\cal L}_{q}$
 and in the present case takes the form
\begin{equation}
{\cal L}_{c} =  {\cal L}_{\alpha'}(c_i) + A e^{\phi} {\cal
L}_{\alpha'}(d_i) + B e^{2\phi} {\cal L}_{\alpha'}(e_i) , \label{efflag}
\end{equation}
where ${\cal L}_{\alpha'}$ is given by Eq.~(\ref{Class}), with the extra 
constant parameters $A$ and $B$ actually controlling the onset of the loop 
corrections. Numerous combinations of these corrections regularise 
the evolution of the background, an example of which is pictured in 
Fig.~\ref{fig}. $A$ and $B$ are typically of order unity and have opposivite 
signs \cite{Cartier:1999vk}. Thus, we can use these non-singular solutions to 
extend different studies based on the tree-level action and probe the 
high-frequency part of the spectrum of relic gravitons.

\section*{Relic gravitons} 
We shall now focus on the generation of gravitational waves arising from 
linearised tensor perturbations. 
Although several mechanisms may contribute to the generation of gravitational
waves from the initial vacuum state of string cosmology (dynamic dimensional
reduction \cite{Gasperini:1992sm}, time-dependence of the dilaton field
\cite{Gasperini:1993dp}), we shall only consider here the usual contribution
arising from the accelerated expansion of the external three-dimensional 
space. As already stressed in a number of papers 
\cite{Gasperini:1993em,Gasperini:1992pa,Gasperini:1993dp}, the production 
of high-frequency gravitons is strongly enhanced compared to that of 
the standard inflationary scenario. Following pionnering work in 
\cite{Gasperini:1997up}, we perturb the full action generated by
Eq.~(\ref{efflag}) around a homogeneous and isotropic solution for the 
scale factor $a(\eta)$ and the dilaton $\phi(\eta)$, where $\eta$ is  
conformal time. In Fourier space and for each normal mode of tensor 
oscillations of our gravi-dilaton background $\psi_{k}$, 
we can rewrite the linearised wave equation in terms of the eigenstates of 
the Laplace operator, $\nabla^2 \psi_{k} = - k^{2}  \psi_{k} $. 
In so doing, we obtain a generalised  wave equation for the tensor 
perturbation:
\begin{equation}\label{psi_full}
\psi_{k}'' + \Bigl\{ k^2[1+c(\eta)] - V(\eta)\Bigr\} \psi_{k} = 0.
\end{equation}
Here a prime denotes differentiating with respect to conformal time. We 
have defined the effective frequency shift 
$c(\eta)\equiv (y/z)^2 - 1$ and the background
quantities including the tree-level $\alpha'$ and loop corrections are:
\begin{eqnarray}
y^{2}(\eta) &=& e^{-\phi} \Bigl[a^2 + \{\oplus\}_{0} + A\{\oplus\}_{1} +
B \{\oplus\}_{2} \Bigr], \label{y_gen} \\
z^{2}(\eta) &=& e^{-\phi}
\Bigl[a^2 + \{\otimes\}_{0} + A\{\otimes\}_{1} + B \{\otimes\}_{2} \Bigr],
\label{z_gen}
\end{eqnarray}
and
\begin{eqnarray}
\{\oplus\}_{n} &=& \alpha' \lambda e^{n\phi} \Bigl\{ 4 c_1 (n-1)^2
{\phi'}^2 + \frac{1}{2} c_2 {\phi'}^2 + 4 c_1 (n-1) \Bigl( \phi''
- \frac{a' \phi'}{a} \Bigr)\Bigr\}, \label{one}\\
\{\otimes\}_{n} &=&
\alpha' \lambda e^{n\phi} \Bigl\{ 4 c_1 (n-1) \frac{a' \phi'}{a}
- \frac{1}{2}c_2 {\phi'}^2\Bigr\}.\label{two}
\end{eqnarray}

The evolution equation for the perturbation Eq.~(\ref{psi_full}) is the main 
result of this study, since it encodes the full contribution 
(through Eq.~(\ref{y_gen})--Eq.~(\ref{two})) 
of those corrections we used to regularise the background evolution.
We stress that in the limit of small coupling and low curvature 
$(\alpha'\rightarrow 0)$ we recover the usual Sch\"{o}dinger-like wave 
equation  since $y,z \rightarrow a e^{-\phi/2}$ and thus 
$c(\eta) \rightarrow 0$.
Indeed, describing an extremely weak coupling and low curvature regime, 
the tree-level solutions of the pre-Big Bang scenario are fully adequate 
in the asymptotic past to describe the gravi-dilaton background 
the metric perturbations are emerging from. This allows us to normalise
our perturbation to positive frequency modes only, 
$\psi_k (\eta) \sim \frac{1}{\sqrt{2k}}e^{-ik\eta}$ for 
$\eta \rightarrow - \infty$.

\begin{figure}[ht]
\begin{center}
\includegraphics[width=5cm,height=4cm]{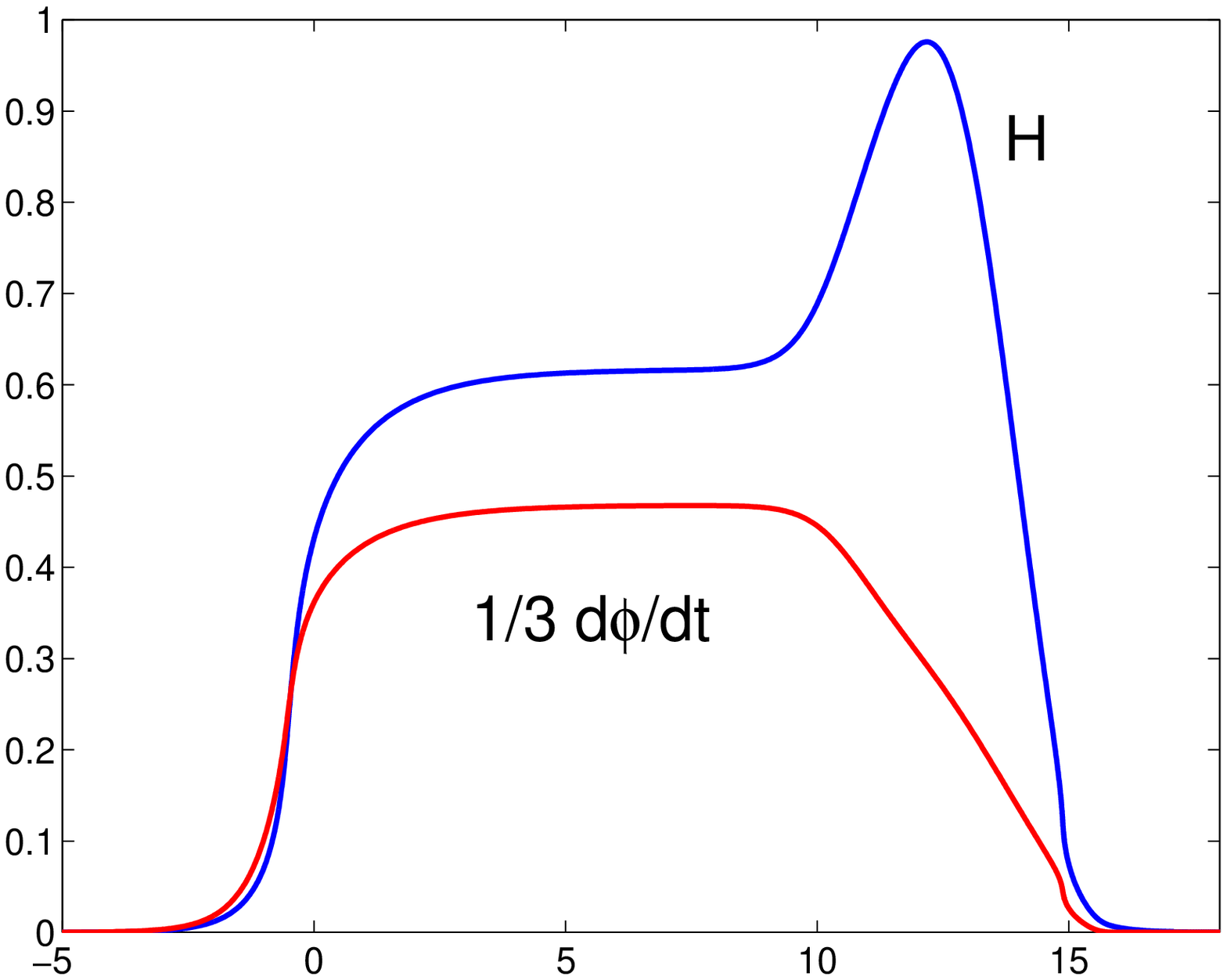}
\includegraphics[width=5cm,height=4cm]{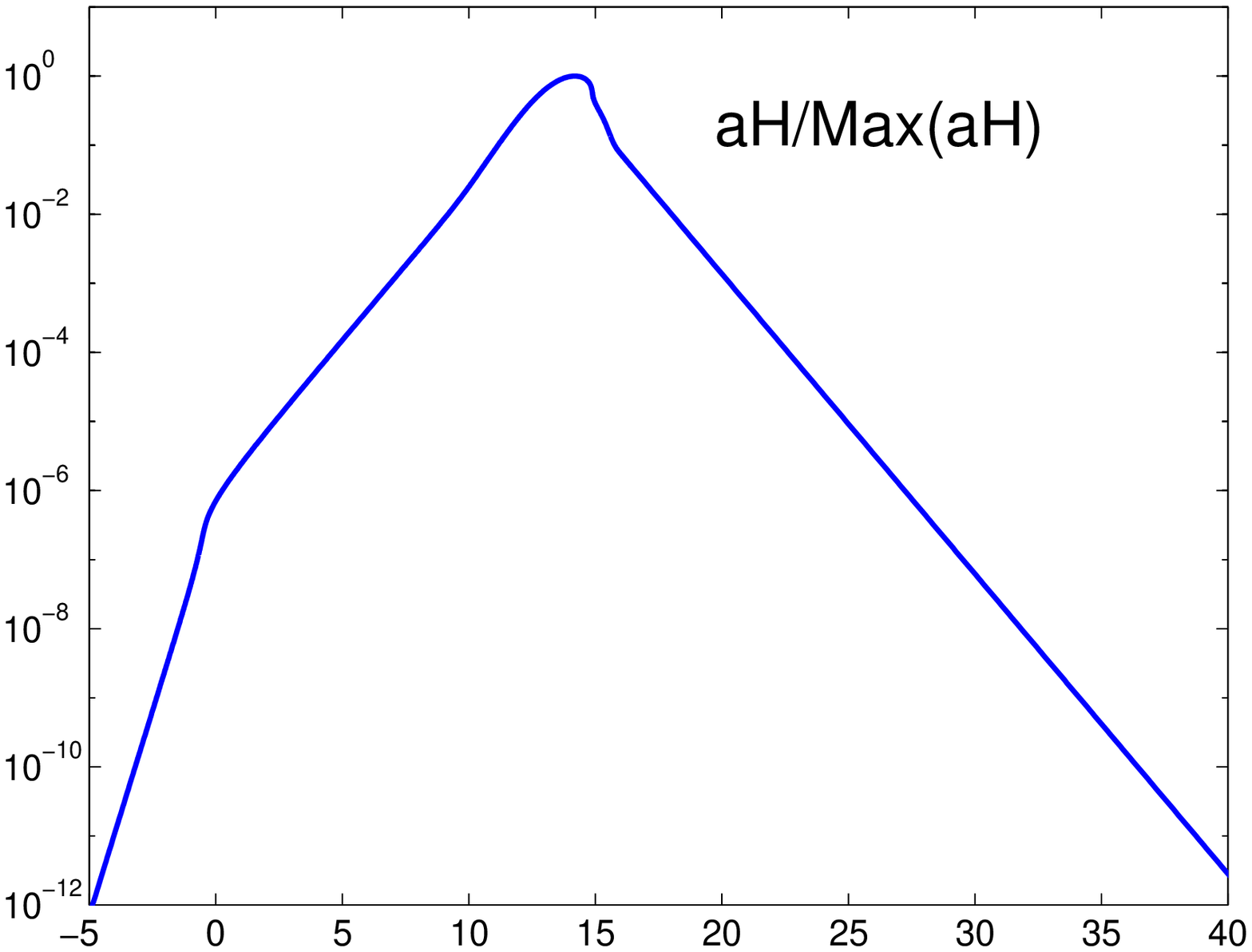}
\includegraphics[width=5cm,height=4cm]{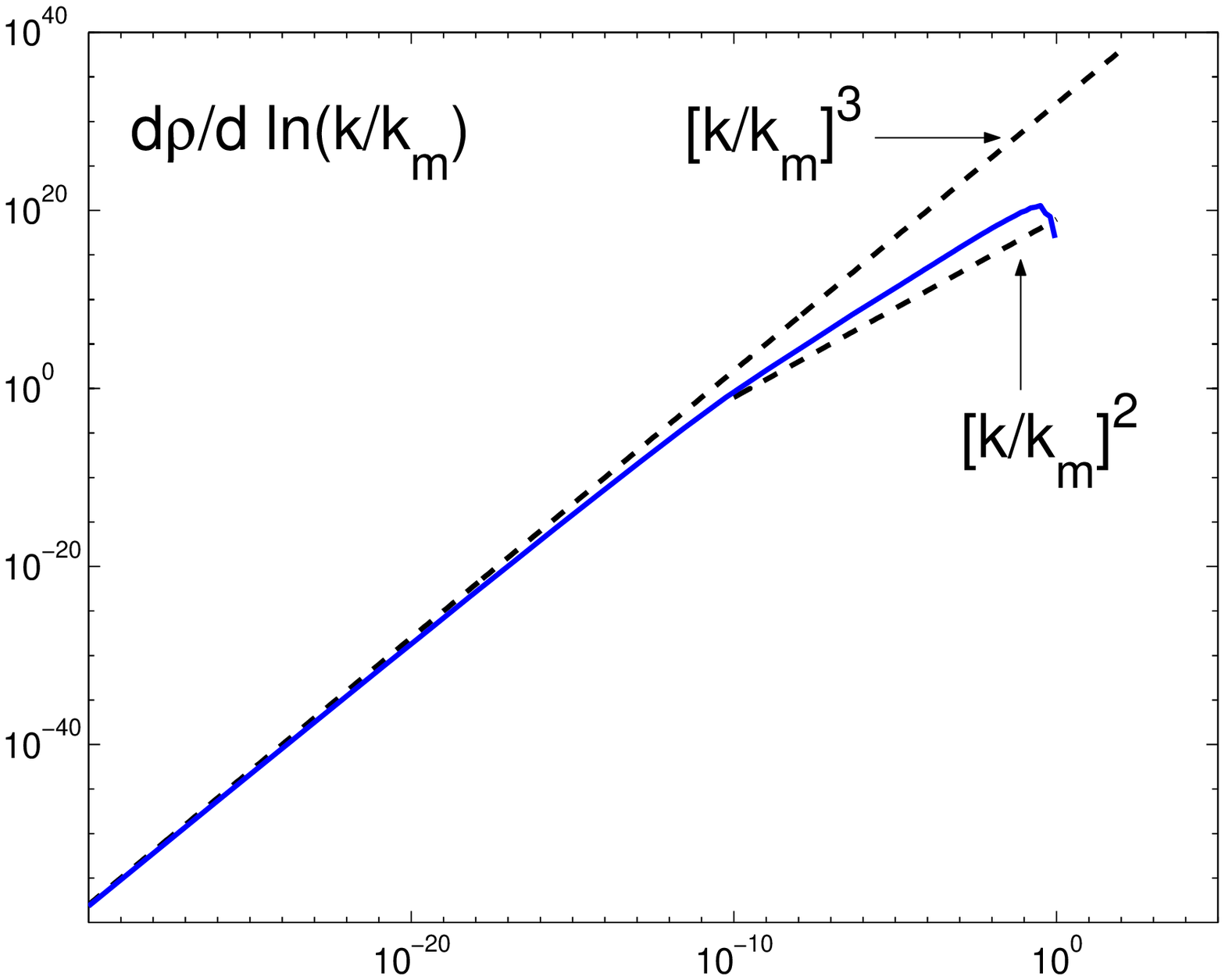}
\caption{
  The left figure shows a non-singular evolution for the Hubble 
  parameter $H=\dot{a}/a$ and for $\dot{\phi}/3$ as a function of the number 
  of e-folds, $N=\ln a$.   
  The centered figure shows the evolution of $aH/{\rm Max}(aH)$ 
  as function of $N$. The pre-Big Bang phase takes place approximatively for 
  $-\infty < N \leq -3$. This inflationary period is followed after a 
  short transition by a string phase with nearly constant Hubble parameter
  and linearly growing (in cosmic time) dilaton for $2 \leq N \leq 9$.
  After a succesfull exit sourced by corrections in the string coupling, 
  the background evolution enters the FRW radiation-dominated phase at 
  $N \simeq 16$. This clearly suggests that comoving modes leaving the
  Hubble radius during the string phase 
  ($10^{-6} \leq k/{\rm Max}(aH) \leq 10^{-2}$)
  will lead to a smaller slope in the spectrum of relic gravitons.
  The right figure shows the corresponding 
  spectral distribution as function of logarithmic interval of 
  frequency, $d\rho_k/d\ln(k)$ in units $k/k_{max}$. 
  The upper dashed line corresponds to the cubic spectrum 
  (pertinent in the small frequency limit) 
  resulting form the dilaton-driven epoch.}
\label{fig}
\end{center}
\end{figure}

The differences between the tree-level wave equation and its generalisation 
Eq.~(\ref{psi_full}) are expected to arise when the curvature scale becomes
large in string units. As a consequence, the low frequency branch of the 
dimensionless spectral distribution of relic gravitons 
remains unaffected by these corrections, and is characterised by a cubic 
slope \cite{Gasperini:1993em,Gasperini:1992pa,Gasperini:1993dp}. 
However, new features due to the corrections arise in the large frequency 
limit. First, a given comoving wavenumber $k$ becomes time-dependent, as 
already observed for a special case in \cite{Gasperini:1997up}. 
Second, the effective potential is strongly reinforced.
As a consequence, such corrections induce an additional amplification 
for comoving modes leaving the horizon during the high-curvature transition.
Figure~\ref{fig} pictures both the evolution of the background quantities and 
the resulting spectral distribution for the relic gravitons.
Of more pertinence is the correspondance between the slope $\xi$ of 
the string branch of the relic gravitons spectrum and the choice of the 
coefficients for the tree-level $\alpha'$ corrections.
Preliminary results \cite{cartier:new}
suggest $1 \leq \xi \leq 3$, which could have considerable implications for 
the detection of such a primordial signal. Indeed, in \cite{Brustein:1997ut},
the authors argue that the frequency peak is typically of order
$\Omega_{gw}(\omega)\simeq 10^{-6}$ for a maximal amplified frequency 
$\omega \simeq 10^{11} [{\rm Hz}]$. 
Saturating this high-frequency end point, and regardless of the duration of 
the string phase with constant Hubble parameter, the spectrum of 
relic gravitons from a pre-Big Bang phase could be at most of 
${\cal O}(10^{-15})$ at $\omega \simeq 10^{2} [{\rm Hz}]$.
The energy density stored in the cosmological gravitational waves 
$\Omega_{gw}(\omega)$ is thus far below the sensitivity 
of the second (planned) generation of spatial detectors!

\section*{Conclusion}
In the context of the pre-Big Bang scenario of string cosmology, 
we have derived a modified tensor perturbation equation including 
those corrections we use to regularise the background evolutions. 
Such modifications, we believe are generic, and lead to new features in 
the string branch of the spectrum of primordial gravitational waves. 
Although the tree-level wave equation is adequate for a first estimate of 
the production of the gravitons during the pre-Big Bang phase, 
this work supports the idea that a full treatment is required when dealing
with the string branch of the spectrum. Preliminary results suggest that,
indepedent of the choice of coefficients in the higher-order corrections,
the slope of the string branch of the spectrum is steeper than first 
expected. As a consequence, the cosmological background of relic gravitons 
may not match with the frequency range probed by (second) planned 
experiments.

\acknowledgments{CC is supported by the Swiss NSF, grant No. 83EU-054774 and 
ORS/1999041014. I am very grateful to E.J.~Copeland and M.~Gasperini for 
stimulating discussions and helpful comments.}

\end{document}